\documentclass[aps,prmaterials,twocolumn,a4,nofootinbib,showkeys]{revtex4-2}

\usepackage[T1]{fontenc}
\usepackage{graphicx}

\newcommand{\ads}{{\mbox{\scriptsize ads}}}

\newcommand{\cut}{{\mbox{\scriptsize cut}}}

\newcommand{\rela}{{\mbox{\scriptsize relax}}}
\newcommand{\start}{{\mbox{\scriptsize start}}}

\makeatletter

\begin{document}

\title{First-principle study of the influence of hydroxyapatite on magnesium surfaces}

\author{Anthony Veit Berg, Ablai Forster, Tim Hansson, Alexandra J. Jernstedt, Emmy Salminen, 
and Elsebeth Schr{\"o}der}\email{Corresponding author; schroder@chalmers.se (E. Schr\"oder)}%
\affiliation{Microtechnology and Nanoscience, MC2,
Chalmers University of Technology, SE-41296 G{\"o}teborg, Sweden}

\date{March 20, 2026}
\begin{abstract}

Hydroxyapatite (HA) on a magnesium (Mg) surface is studied using density functional theory, 
to help understand the effect of HA coating and alloying in the surfaces of Mg-based biodegradable implants.
We determine the adsorption energies and structural changes of a single layer of HA on pure Mg(0001) and on sparsely calcium (Ca) or zinc (Zn) doped Mg(0001) and find that both Zn and Ca doping improves the adsorption, except in a few positions of HA relative
to the dopant position. 
All adsorption configurations, whether with pure or doped Mg surfaces, show deformation of the surface and HA layer.
For Ca doping, we found that for a certain adsorption configuration, the dopant Ca atom moves out of the Mg surface and into the 
HA layer, leaving behind a Mg vacancy in the top layer of the Mg surface. 
Plots of electron density changes show that electrons accumulate around the Ca dopant and the neighboring Mg atoms, while in
Zn doping this is less pronounced.
Overall, our results demonstrate that the dopant choice and relative position of HA influence the interaction between HA and Mg-surfaces, and affect
both adsorption energies and atomic and electronic structures.
\end{abstract}

\keywords{density functional theory, magnesium, hydroxyapatite, vdW-DF-cx
}
\maketitle

\hyphenation{over-estimated mole-cules}


\section{Introduction}

Orthopedic implants are used to stabilize fractures and replace damaged tissue. In practice, metals like titanium, cobalt chromium alloys and stainless steel are often used because they tolerate high loads and wear \cite{Kim_2020}. 
Magnesium (Mg) has also gained interest for orthopedic implants due to its biocompatibility, mechanical compatibility with bone and since it naturally degrades within the human body \cite{Tsakiris_2021}. 
Many alloys currently used for implants are much stiffer than bone, causing the implant to absorb mechanical load. 
As a result, stress shielding slows down the healing process \cite{Niinomi}. 
The stiffness of Mg, on the other hand, resembles that of natural bone better \cite{Luthringer2014}.

Mg corrodes easily due to its low electrochemical potential. 
This leads to a release of Mg ions, which are naturally present in the body and are thought to support bone healing \cite{Zhao2016,Han2019}. 
Titanium implant materials, such as Ti-6Al-4V, may pose safety concerns with long-term release of 
vanadium or aluminum since these elements could be a probable cause for osteomalacia, peripheral neuropathy and Alzheimer's disease \cite{Lin2005}. 
For Mg the main concern is the fast corrosion rate of pure Mg, which may lead to H$_2$ and possibly other gas cavities that may damage nearby tissue, and in severe cases cause necrosis \cite{Tsakiris_2021,Noviana2016}, although in some studies the gas is found to mostly modify the local environment
rather than being harmful per se \cite{amara25}. 
As degradation progresses, the implant may fragment under physiological compressive loads. This damage can propagate as penetrating cracks, resulting in a loss of mechanical integrity \cite{Guo2025}.  
 
Given the challenges associated with Mg degradation, many Mg-based alloys and coatings have emerged as promising approaches to help implants last longer and support bone regeneration. 
For example, Mg alloyed with small amounts of Zn and Ca shows good biocompatibility 
\cite{Mohamed2019_Mg08Ca_JMagnesAlloys}, and
after immersion in simulated body fluid such surfaces show sign of forming magnesium hydroxide (Mg(OH)$_2$) and hydroxyapatite (HA) \cite{petrovic20,willumeit11}.
Further, HA has has a tendency to adhere to Mg-based substrates, indicating its potential to serve as a protective layer that mitigates the degradation of an implant \cite{yang2015nano,iskandar2013,kim2013,zhang2008,kannan2015,dunne2024}. 

In this study, density functional theory (DFT) calculations are used to examine the stability of HA coatings on Mg-based surfaces, 
including the effects of sparse zinc (Zn) or calcium (Ca) alloying of the Mg surface. 
We determine adsorption energies and structures, and study the changes in electron distribution that arise from the adsorption. 
Our results provide insight into factors influencing the adsorption and stability of HA on these surfaces. 

The text is organized as follows: Section II describes the materials while Section III outlines the methods. 
Section IV presents and discusses the results, and Section V contains our conclusions and a summary.


\section{Materials}

The choice of implant material depends on many factors.
It is often desirable for implant materials to have a stiffness (Young's modulus) that is comparable to bone to avoid stress shielding. The elastic modulus in the bone, depending on the site and bone type, is roughly 4-30 GPa \cite{Geetha2009}. 
Mg has a Young's modulus of approximately 44 GPa and resembles the stiffness of natural bone better than other commonly used metals \cite{Luthringer2014}. 
By comparison, stainless steel and Co-Cr have a Young's moduli around 200-240 GPa and the widely used Ti-6Al-4V near 110 GPa \cite{AbdElaziem2024}. 
If the implant cannot stay permanently in the body (e.g., growing bone in children), a further drawback of stainless steel and Ti-6Al-4V  is that a second surgery may
be needed for removal \cite{Dehghan2024}, while Mg degrades and the challenge there is to have it degrade at 
a pace corresponding to the healing of the bone. Alloying and coating are methods to inhibit the corrosion.

\begin{figure}[tb]
\centering
\includegraphics[width=\columnwidth]{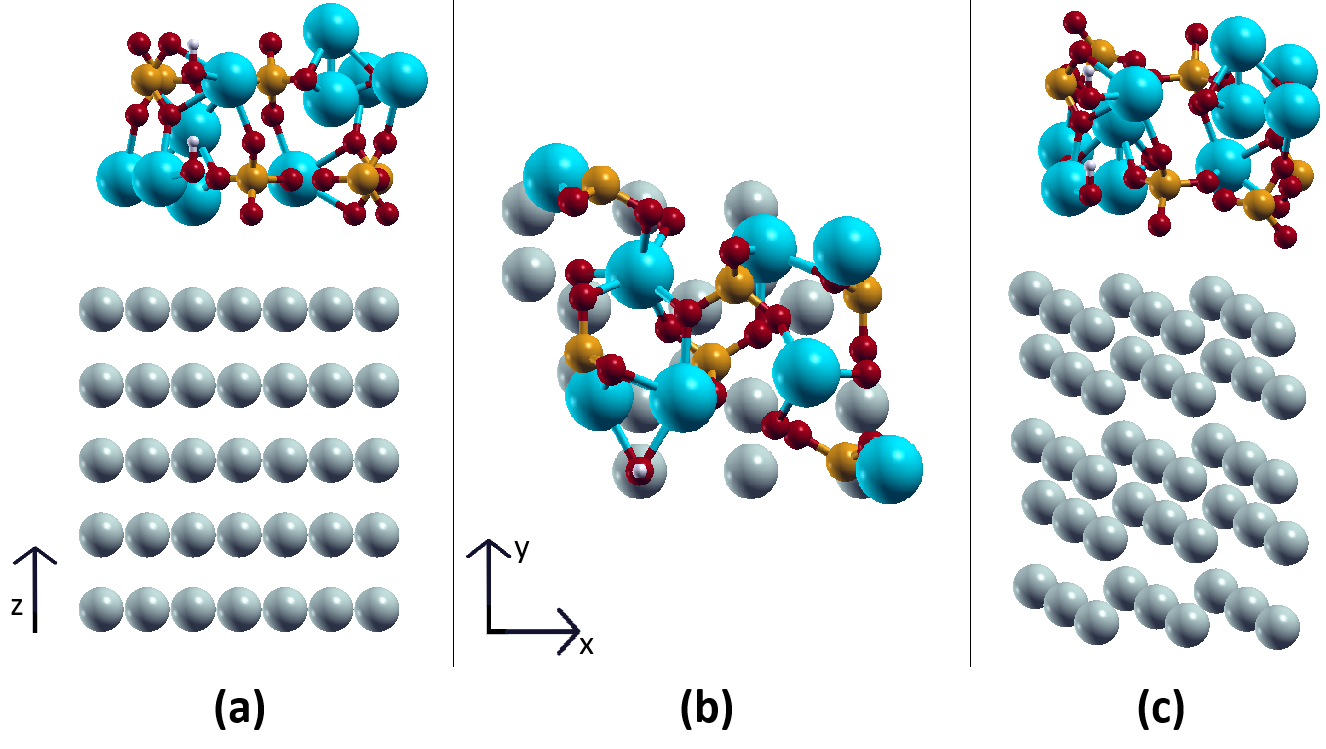}
\caption{\label{fig:HA_mg_olika_vinklar} Hydroxyapatite (HA) on the Mg(0001) surface, shown from three perspectives to illustrate the structural arrangement and relative positioning. (a) Side view ($x$-$z$ plane). (b) Top view ($x$-$y$ plane). (c) Angled view.
Color coding of atoms: magnesium - grey, oxygen - red, phosphorus - orange, calcium - blue, hydrogen - white.
All structures and electron densities visualized with XCrySDen.}
\end{figure}

For modelling coating we here use a layer of HA on top of the Mg(0001) surface, either on pristine Mg(0001) or with sparse doping of Zn or Ca. 
Figure \ref{fig:HA_mg_olika_vinklar} provides an overview of the combined structure of HA on Mg(0001) as seen from the side, 
the top and a tilted side view, with the cartesian coordinate system used later indicated in (a) and (b). 
The figure shows an isolated HA layer rigidly put on top of pure Mg(0001), unaffected by the 
presence of the HA layer and before any deformations of Mg(0001) or HA have been taken into account. 
All structures and electron densities are visualized with XCrySDen \cite{Kokalj2003}. 

\subsection{Mg(0001) with alloys}

The crystalline structure of Mg bulk at ambient conditions is the hexagonal close-packed (hcp) structure.
With the methods used here, the lattice parameters are found \cite{schroderPP} to be
$a= 3.192$~{\AA} and $c= 5.186$ {\AA}. 
We study the basal plane Mg(0001) which is the most stable direction and which is commonly used for surface calculations in DFT studies 
due to its simple, unreconstructed structure \cite{ScFaKi04,Li2009,Li2010,Nezafati2016,fang18,Han2021,Xing24,zhe25}, 
while other low-index directions such as the prismatic ($10\bar{1}0$) and pyramidal ($10\bar{1}1$)
have higher reactivity \cite{Fiesinger2022,Ji2016}. 
For its simplicity and compatible surface unit cell size relative to HA, Mg(0001) is chosen here as a model structure to investigate interactions with HA. 

Surface doping of Mg(0001) with Zn and Ca is of particular interest due to their compatibility with Mg and 
their natural occurrence in the human body. 
Zn is an essential mineral which has vital functions in the body, some being catalytic activity of enzymes, protein- and DNA synthesis \cite{ODS_Zinc_HealthProfessional}. 
The crystalline structure of Zn, like Mg, is hcp, making Zn compatible with the structure of
Mg(0001) \cite{Romzi2022_EffectZn_on_Mg}. 
Studies have shown that corrosion rates of Mg decreases with Zn alloying and the rate depends on the Zn content 
\cite{BakhsheshiRad2014_MgZn_Biocorrosion,Xu2022_CorrosionMgAlloys}. 

Ca, on the other hand, is the most abundant mineral in the body \cite{ODS_Calcium_HealthProfessional}. 
It is an essential mineral (like Zn) and a main component of HA found in bones and teeth. 
The corrosion behavior can improve with Mg-Ca alloys, an the corrosion rate depending on the Ca content 
\cite{Wan2008_MgCa_MaterialsDesign}. 
Ref.\ \cite{Mohamed2019_Mg08Ca_JMagnesAlloys} reported that an Mg-Ca alloy implant immersed in Hank's balanced salt solution showed excellent 
biocompatibility, with formation of HA on the surface, while Ref.\ \cite{Li2008_MgCa_Biomaterials} reported HA formation on the 
surface of a Mg-Ca alloy with 1 weight percent (wt\%) of Ca during both in vitro and in vivo corrosion. 

\subsection{Hydroxyapatite}

HA  is a calcium phosphate that is one of the main components of human bone \cite{Dorozhkin2007-CalciumOrthophosphates}. 
It is widely recognized for its osteoconductivity and biocompability and shows great potential to be used in areas like bone tissue engineering 
\cite{Shi2021-HApReview,Ielo2022-HApBiocomposites}. 
Due to its brittleness and low fatigue resistance, HA is not suitable as a load-bearing implant and it is primarily used in non-load-bearing applications \cite{Prakasam2015DenseHA,Habibah2022-HydroxyapatiteDentalMaterial,Kauke-Navarro2024-HApFacialImplantology,mondal2023hydroxyapatite,zhang2022}.
However, applying HA as a surface coating on metals can affect the corrosion behavior and the  properties of the material, which could be useful in implant applications \cite{kannan2015,Asri2017-CorrosionSurfaceModificationMetals,Tomozawa2010SCT}.

\section{Methods}

\subsection{Density functional theory modelling of Mg(0001)} 

Our results are based on DFT calculations with 
the van der Waals-inclusive functional vdW-DF-cx \cite{dion04p246401,berland14p035412,berland14jcp,schroder17chapter}, using the
plane wave code \texttt{pw.x} for the DFT calculations and \texttt{pp.x} for postprocessing, both codes from 
the Quantum ESPRESSO suite \cite{QE,espresso,QE_2}. 
We use PAW-based pseudopotentials from PSLibrary \cite{pslibrary,dalcorso14} to describe the atoms, except for Mg, 
which is described by a PAW pseudopotential that one of us created and thoroughly tested in Ref.\ \cite{schroderPP}.

The Mg(0001) surface is modeled by slabs with periodic boundary conditions,
in surface unit cells that consist of $3\times 3$ hcp bulk unit cells laterally and 5 layers of Mg as thickness of the slab. 
This lateral size is chosen as it approximately fits with one unit cell of HA.
All surface calculations have vacuum added on top, 
in order to avoid 
interactions with repeated slabs along the $z$-direction.
This leads to a total cell size of 36 {\AA} in height, 
leaving around 16 {\AA} of vacuum above the slab when HA is added on Mg(0001). 
The bottom three Mg layers are kept fixed in the atomic positions obtained from a calculation
with a fully relaxed slab of 23 layers \cite{schroderPP,bolin26}, while 
 all other atoms are allowed to move according to
the Hellmann-Feynman forces (except when stated).

\subsection{Convergence tests and computational parameters}

Aiming for a calculational accuracy of at least 10 meV per supercell we carried out 
convergence studies of the number of Monkhorst-Pack \cite{monkhorst76p5188} k-points 
needed in the supercell with HA included, 
as well as the plane wave energy cutoff for the kinetic energy, $E_\cut$ and electron density $E_\cut^\rho$. 
The k-point convergence test was conducted using $n_k \times n_k \times 1$ grids with
$n_k=2$, 4, 6, 8, 10, and 12. 
For each mesh, a standard relaxed HA-on-Mg(0001) structure was calculated. 
Then, the HA layer was uniformly moved 0.1 {\AA} away from Mg(0001)
and a new calculation with fixed atom positions was performed on the structure. 
From the difference in total energy between these two configurations  we found that
the best balance between calculation accuracy and computational cost
was obtained with the $6\times 6\times 1$ k-point mesh.

In a similar series of test calculations for 
energy cutoffs of the plane-wave basis, $E_\cut$, and electron density, $E_\cut^\rho$,
we tested values 40/320, 50/400, 60/480 and 70/560 Ry, using the $6\times 6\times 1$ k-point mesh,
and found the optimal values $E_\cut=50$ Ry and $E_\cut^\rho=400$ Ry. 
These are the values we use throughout this study.

To report directional changes in atomic positions and changes in atom-atom distances we introduce a cartesian coordinate system 
as indicated in Figure \ref{fig:HA_mg_olika_vinklar}, with $x$- and $y$-directions in the Mg surface (0001) plane and $z$ in the [0001] direction. 

\subsection{Adsorption energy}

The adsorption energy is calculated as the difference between the total energy of the full system and the energies of the isolated components,
\begin{equation}
E_{\mathrm{ads}} = E_{\mathrm{tot}} - \left( E_{\mathrm{HA}} + E_{\mathrm{Mg}} \right) \,.
\end{equation}
Here, $E_{\mathrm{tot}}$ is the total energy of HA adsorbed on the Mg surface after full structural relaxation. 
The surface reference energy $E_{\mathrm{Mg}}$ is calculated for the clean or doped Mg surface, depending on the system. 
The reference energy of HA, $E_{\mathrm{HA}}$ is obtained at the lateral lattice constants that fit the Mg(0001) surface and
allowing the atomic positions to relax.

\subsection{Hydroxyapatite}

HA has the stoichiometric formula C$_{10}$(PO$_4$)$_6$(OH)$_2$ in a 
hexagonal bulk unit cell (space group P$6_3$/m), with
44 atoms per unit cell. 
We obtain the structure from Materials Project \cite{materialsproject} but further
relax the lattice 
constants and atomic positions using the vdW-DF-cx functional. 
Bulk HA has a layered structure, for which we find in-plane (vertical) lattice constant $a=9.41$ {\AA} 
($c=6.87$ {\AA}), in excellent agreement with values from experiment \cite{posner58} $a=9.43$ {\AA} and $c=6.88$ {\AA}.
The in-plane lattice constant is approximately three times that of Mg(0001), 
for HA coating on Mg we therefore use one layer of HA on $3\times3$ Mg(0001), 
thus stretching HA approximately 2\% to fit the underlying Mg surface.

\begin{figure}[tb]
\centering
\includegraphics[width=\columnwidth]{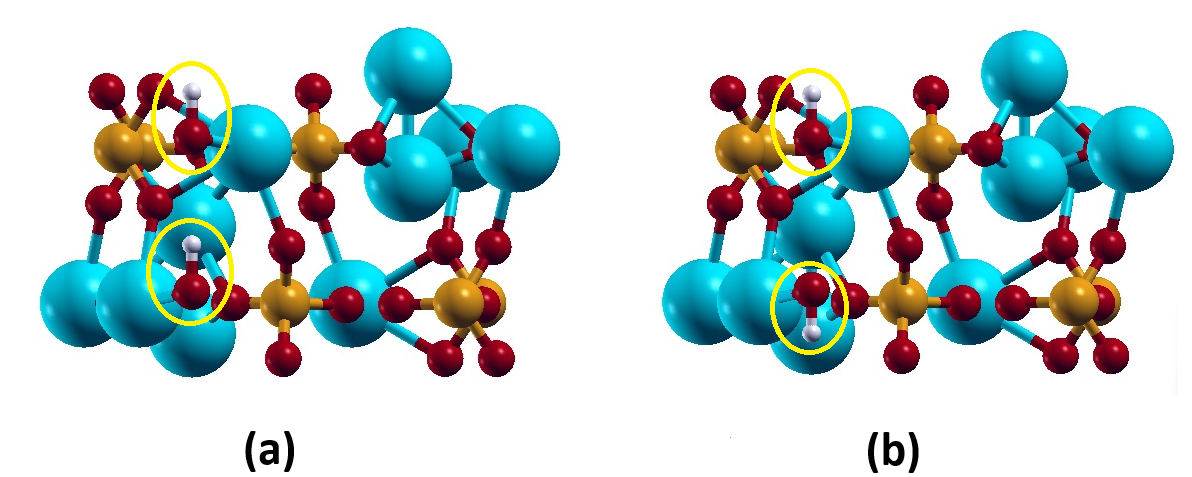}
\caption{\label{fig:HA_dipol_byte.png} HA structures with OH group orientations (a) as energetically favoured for a single layer of HA (b) as in bulk HA.
The two OH groups in each panel are marked with yellow circles. 
Atom colors as in Figure \protect\ref{fig:HA_mg_olika_vinklar}.
}
\end{figure}

The two hydroxyl groups in a HA layer in bulk align along the $c$-axis, 
Figure \ref{fig:HA_dipol_byte.png}(a), creating a net dipole. 
However, as we describe in the Results section, when a single layer of HA is isolated, 
it is energetically favourable to have both H atoms pointing outward, thus flipping one OH group, 
as illustrated in Figure \ref{fig:HA_dipol_byte.png}(b). 
We therefore use this structure for the coating adsorption studies presented here.
 
\subsection{Zinc or calcium alloying of Mg(0001)}

To investigate the effect of sparse Zn or Ca alloying of Mg(0001), one Mg atom 
in the surface top layer of Mg(0001) is substituted in the $3 \times 3$ supercell. 
This corresponds to a Zn (Ca) concentration of 5.8 wt\% (3.6 wt\%) of the atoms in the two exposed 
layers of Mg. 
Several structures were constructed with varied dopant positions in the top layers, relative to the HA coating, 
in order to explore how the position of the alloying atom may influence the interaction. 
The positioning of the substitutes was determined based on the optimal placement of HA over pristine Mg(0001). 
Each dopant position was selected to examine interaction with a given HA atom or with no HA atom specifically targeted. 

\begin{figure}[tb]
\centering
\includegraphics[width=\columnwidth]{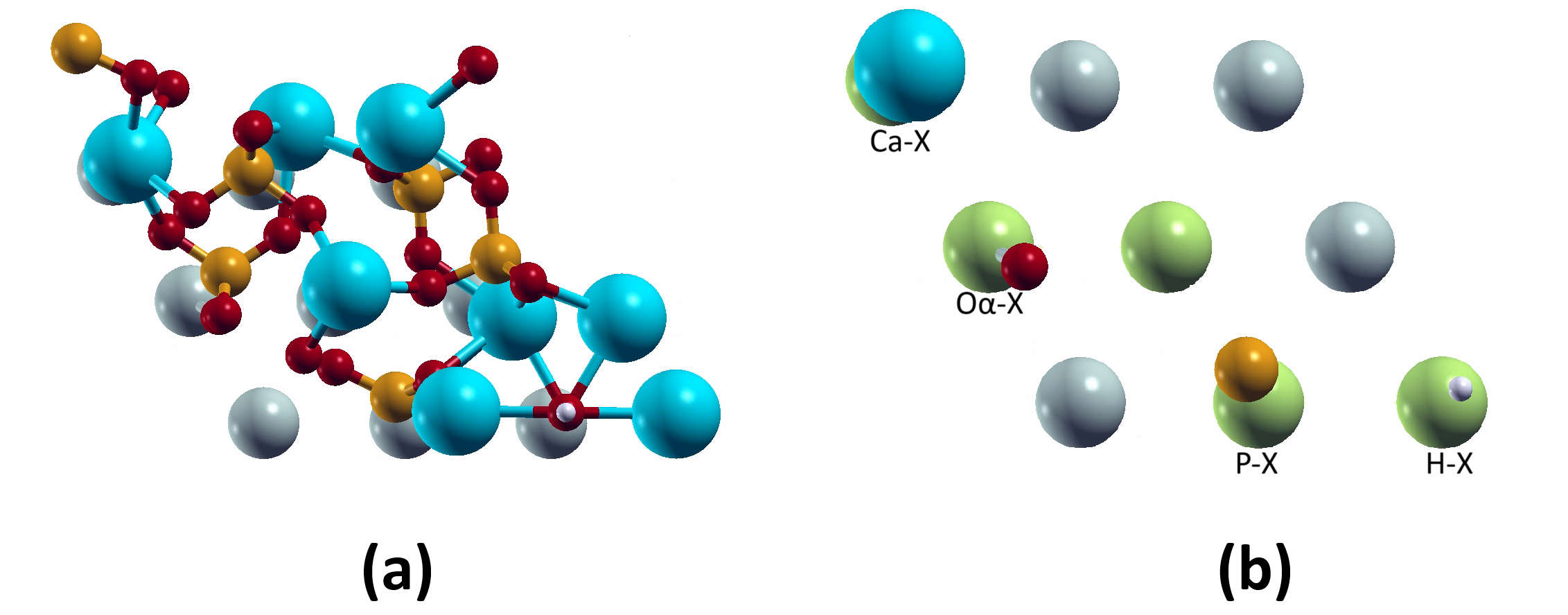}
\caption{\label{fig:Sub_atoms}Atomic configurations showing the positioning of the substituted atoms. 
(a) All of HA above the top layer of Mg. (b) Substituted atom positions in green and including only the atoms in HA that are directly above them. Notation Ca-X means HA-bound Ca atom over substituted atom X (Zn/Ca). Only one Mg atom at a time is substituted.
Other atom colors as in Figure \protect\ref{fig:HA_mg_olika_vinklar}.}
\end{figure}

The position of each Mg top layer atom within the $3\times 3$ supercell is designated by an index notation ($n$, $m$), where (0,0) corresponds to the lowest left-hand corner of the supercell, Figure \ref{fig:Sub_atoms}. 
The selected sites for substitution, Fig.\ \ref{fig:Sub_atoms}(b), are position (2,0), which is located beneath a hydrogen (H) atom of a -OH group, 
position (0,2) underneath a Ca atom, position (0,1),
located close to an oxygen (O) atom, position (1,1) at the center of the structure
and not under any particular HA atom, and finally position (1,0), which is located underneath a phosphorus (P) atom.

\section{Results and discussion}

\subsection{Hydroxyapatite on pristine Mg(0001)}

HA in the bulk structure has both OH groups pointing in the same direction, 
Figure \ref{fig:HA_dipol_byte.png}(a), but for a single layer of HA we test and find that a structure with both H atoms pointing out of the layer, 
Fig.\ \ref{fig:HA_dipol_byte.png}(b), is energetically favored:
the gain in energy for the H-out structure is 9.9 meV/{\AA}$^2$, or 0.79 eV per flipped OH group, compared to 
the structure with both H atoms pointing the same direction. 
We also test the
structure with both H atoms pointing into the layer, and find that the cost, again
compared to having the H atoms in the same direction, is 7.0 meV/{\AA}$^2$, 
or 0.56 eV per flipped OH group. 
We therefore use the structure with H-atoms out, Fig.\ \ref{fig:HA_dipol_byte.png}(b),
for our analysis of a single HA layer as coating on Mg(0001). 

The optimal positioning of the HA coating layer relative to the Mg surface was investigated by systematically translating the 
HA layer across the Mg surface.  
Within one unit cell of Mg(0001) a total of 25 equally distributed positions were evaluated.
In each such position we allowed all atoms of the HA layer to relax in the direction perpendicular to the surface, 
to keep the HA layer from ``sliding'' off unfavorable positions,
and the atoms two top layers of the Mg surface were allowed to move in all directions.
This approach enables identification of the energetically most favorable adsorption site by sampling a representative range 
of positions across the Mg surface.  

\begin{figure}[tb]
\centering
\includegraphics[width=\columnwidth]{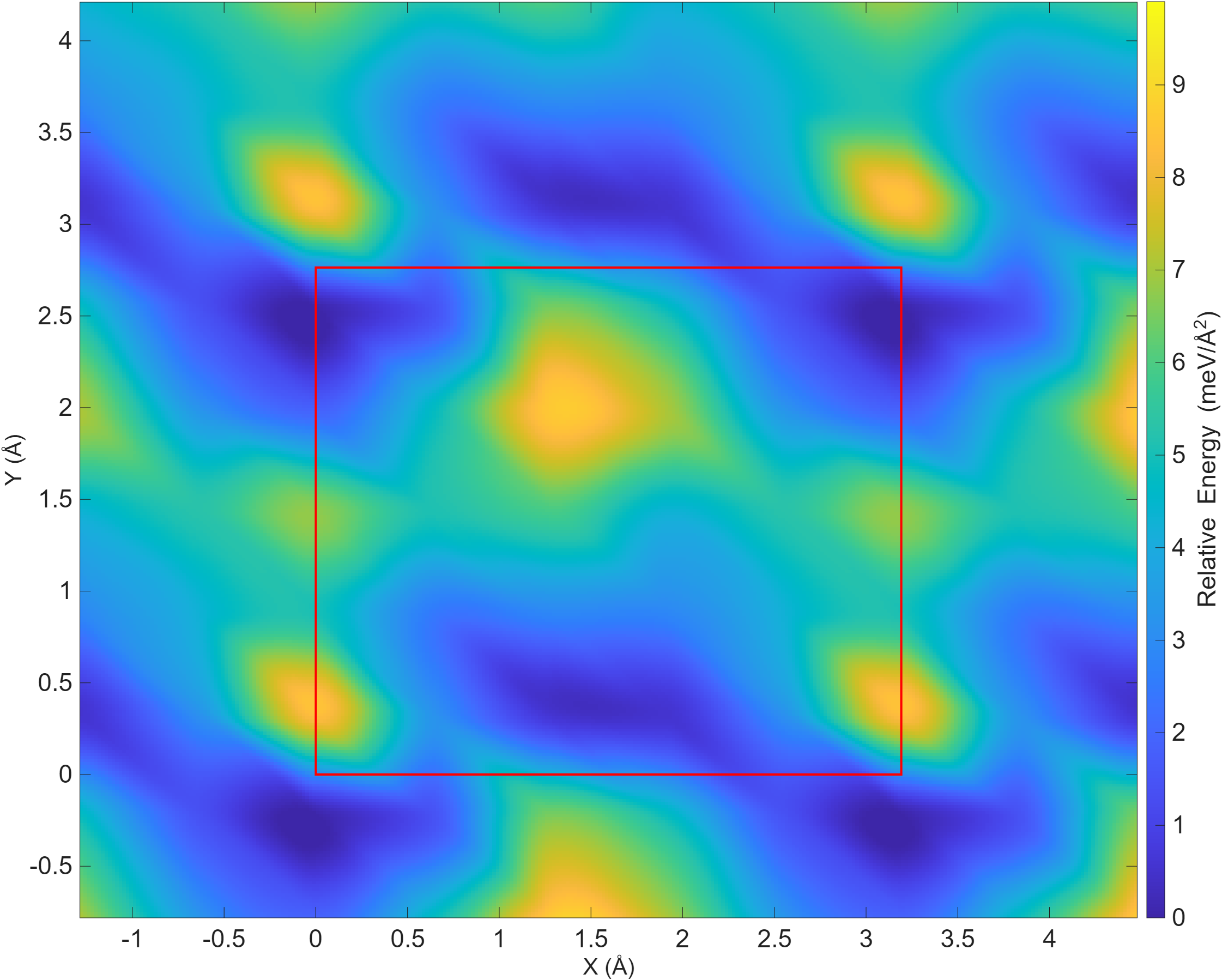}
\caption{\label{fig:contour_graph} Potential energy contour plot for relative position of HA on Mg(0001), for estimating the optimal positioning of HA over the Mg surface. Energies relative to the energy of the optimal position.}
\end{figure}

Figure \ref{fig:contour_graph} shows the PES based on the 25 calculated positions with restraints on HA atomic lateral positions. 
The region calculated is highlighted with a red rectangle, values outside the rectangle are periodic copies. 
The plot indicates that there is a preferred adsorption site for HA over the Mg surface. 
However, the calculated energy differences between the optimal and the energetically worst position is relatively small, 
at 9.9 meV/{\AA}$^2$, which is only double the energy required to slide a graphene layer 
on a surface of graphite \cite{naurin26}. 
This suggests that the corrugation of HA on Mg(0001) is small, as also seen for a layer of
magnesiumhydroxide (Mg(OH)$_2$) on Mg(0001) \cite{bolin26}
and discussed more for DFT-calculations of chloroform on graphene in 
Ref.\ \cite{akesson12p174702}.
A HA layer may thus slide across the Mg surface rather easily, for example, if a small force is applied or if the Mg surface develops cracks or irregularities. 
While this mobility might help release stress between the interface of HA and Mg, it could also make the coating less stable.

After further relaxation of all atoms at the optimal adsorption site, 
the system converges to $E_\ads = -14.4$ meV/\AA$^2$ for HA on pristine Mg(0001).

\begin{figure}[tb]
\begin{center}
\includegraphics[width=\columnwidth]{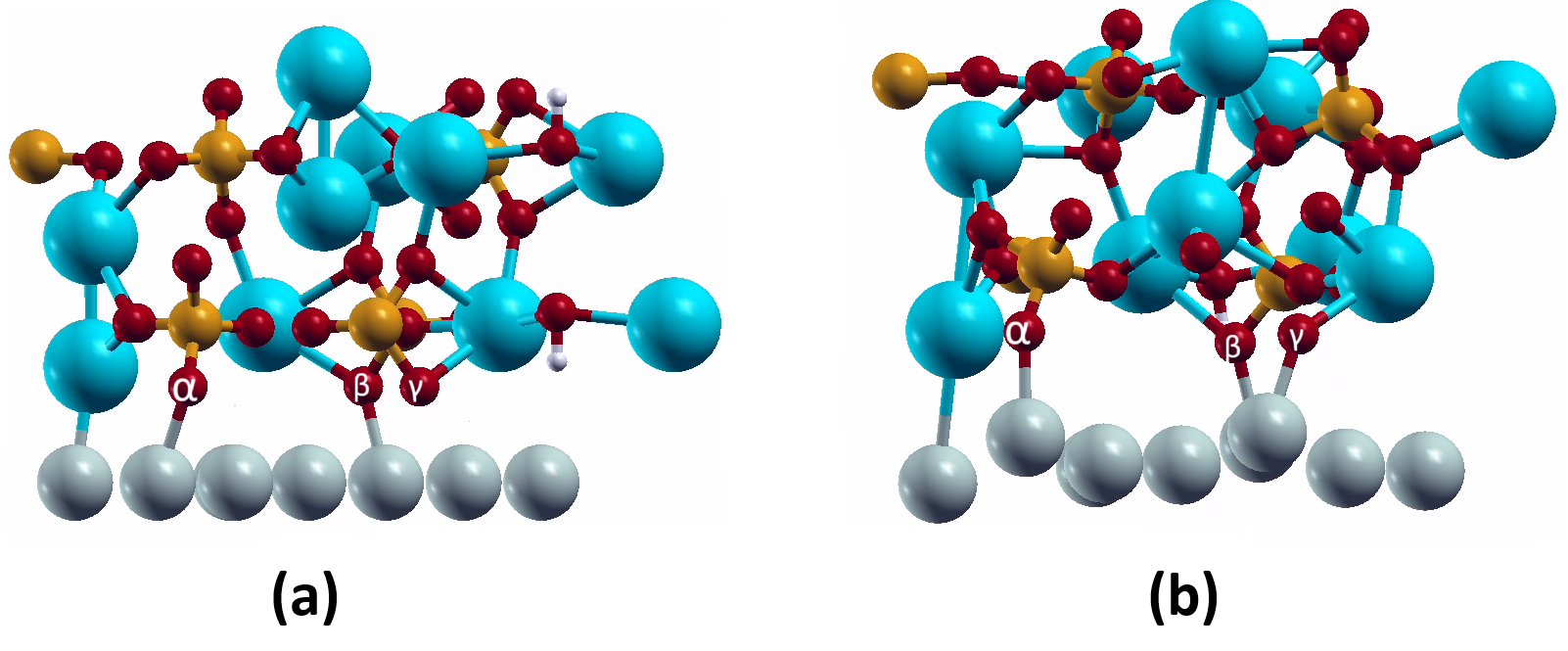}
\end{center}
\caption{\label{fig:HA_before_after}Structural changes of HA in the optimal translational position, starting from (a) the structure of an isolated layer of HA (slightly stretched in the lateral directions) and clean Mg(0001) brought together at 2 {\AA} separation, and (b) after optimization of the atomic positions.
Atom colors as in Figure~\protect\ref{fig:HA_mg_olika_vinklar}.
}
\end{figure}

\begin{table}[tb]
    \centering
    \caption{Changes in distances between pairs of atoms in the alloyed Mg surface and HA,
    relative to the structure of HA on pristine Mg(0001) shown in Figure \protect\ref{fig:HA_before_after}(b).
    The changes are given in $\Delta x$, $\Delta y$ and $\Delta z$, and 
    the change in the length of the distance vector, 
    $\Delta r=|\mathbf{R}_\rela|-|\mathbf{R}_\start|$. 
    Positive (negative) values signify that the distance in that direction between the two atoms has increased (decreased). 
    Also shown is the distance between atoms after
    the relaxation, $R_\rela=|\mathbf{R}_\rela|$.
    Values are for for pure Mg(0001) (top of table), for alloying with Ca (middle part of table), and for alloying with Zn
    (lower part of table).
    For the latter two the alloying atom is placed beneath the HA atom mentioned.
    All values in {\AA}.
    }
    
    \begin{tabular}{ll|ccccc}
    \hline
      Doping & Atom pair & $\Delta x$ & $\Delta y$ & $\Delta z$ & $\Delta r$ & $R_\rela$\\
        \hline
       & Ca-Mg* & - & - & - & - & 3.085 \\
      &  O$_\alpha$-Mg* & - & - & - & - & 2.088 \\
       & O$_\beta$-Mg* & - & - & - & - & 2.032 \\
     &   O$_\gamma$-Mg* & - & - & - & - & 2.053 \\
        \hline
     (0,2) &   Ca-Ca & 0.281 & 0.297 & $-$0.148 & $-$0.102 & 3.186 \\
     (0,1) &    O$_\alpha$-Ca & $-$0.118 & $-$0.039 & $-$0.138 & $-$0.154 & 2.253 \\
     (1,0) &   P-Ca & 0.050 & $-$0.067 & $-$0.137 & 0.139 & 3.439 \\
     (2,0) &   H-Ca & 0.406 & $-$1.341 & 0.852 & 0.409 & 2.484 \\
        \hline
     (0,2) &   Ca-Zn & $-$0.004 & 0.152 & 0.140 & 0.156 & 2.929 \\
     (0,1) &   O$_\alpha$-Zn & $-$2.278 & 0.073 & $-$0.637 & $-$1.598 & 3.698 \\
     (1,0) &   P-Zn & $-$0.217 & $-$1.246 & $-$0.682 & $-$0.912 & 4.212 \\
     (2,0) &   H-Zn & 0.117 & $-$0.037 & $-$0.018 & 0.010 & 2.883 \\
        \hline
    \end{tabular}
    \label{tab:distance_atoms}
\end{table}

As shown in Fig.\ \ref{fig:HA_before_after}, the Mg atom positioned beneath the lowest Ca atom of HA moves slightly downward, away from HA. 
Meanwhile, three Mg atoms situated beneath or near the lowest O atoms in HA (marked by $\alpha, \beta$ and $\gamma$) move towards the O atoms.  
The top part of Table \ref{tab:distance_atoms} summarizes the distances between these atoms and the middle part of 
Table \ref{tab:zpos_combined} (line HA-Mg) summarizes the vertical changes of the Mg atoms relative to their positions 
before HA adsorption: the Mg atoms in positions (1,0) and (0,1) move outwards, toward the lowest HA O atoms, and inwards 
in positions (0,2), away from the lowest HA Ca atom. Other Mg atoms do not move much. 

All three O-Mg pairs, listed in the top part of Table \ref{tab:distance_atoms}, end up with a similar total separation $R_\rela$, 
at about 2.05 {\AA}, which corresponds well with the out-of-plane Mg-O separation (2.005 {\AA}) in the wurtzite (WZ) phase of MgO. 
WZ-MgO is the first (metastable) phase that appears in the top of Mg(0001) as 
oxidation starts \cite{zhe25}, while the final, stable rock salt phase has a slightly longer Mg-O separation.
Meanwhile, the Ca-Mg pair shows a separation of about 3 {\AA}.
These displacements show that the surface atoms of Mg are influenced by both the Ca atom and the O atoms of the HA layer, 
leading to a rearrangement in the surface as seen in Fig.\ \ref{fig:HA_mg_olika_vinklar}.    

As mentioned in the previous section, we stretch the HA layer from the bulk lateral size to fit the $3\times 3$ Mg(0001) surface when adsorbed.
When reporting the adsorption energies as a reference we use
a free-standing HA layer with the same lateral lattice constant as $3\times3$ Mg(0001) and with relaxed atomic positions.
If we instead were to use either a layer of HA in its bulk (at lateral lattice constant $a=9.33$ {\AA}) 
or a freestanding layer of HA (with the even smaller
optimized lattice constant $a=9.26$ {\AA}) the reported adsorption energies would include the cost of ``cutting'' 
out the HA layer from bulk and stretching it (at a cost 
62.2 meV/{\AA}$^2$, not counting the gain of flipping one OH group once isolated from bulk), or the cost of stretching the free-standing layer from
its smaller optimized lattice constant (cost 10.3 meV/{\AA}$^2$). 
While we do not explicitly include these energy costs, they can be added to all adsorption energies provided in this study if of relevance.
Given the smaller optimal lattice constant for the freestanding HA layer and the relative ease with which HA can slide on the Mg surface,
there is a possibility that with large patches of HA on the surface, the HA structure will shrinks in size in relation to the Mg surface, 
potentially creating holes or cracks in the HA layer.

\subsection{Alloying of Mg(0001)}

When one of the Mg atoms is substituted by a Zn or Ca atom the positions of the atoms in the surface change.
For Zn alloying, Zn dips into the surface, and Ca moves partly out of the surface, as
already seen in Ref.\ \cite{bolin26} for the same calculational conditions, but in a $5\times 5$
surface unit cell, while we here use a $3\times 3$ surface cell.
For the present surface unit cell, the 
upper part of Table \ref{tab:zpos_combined} shows the displacement along the $z$-direction of the dopant atom compared to the 
position of the Mg atom it replaces. 

In this smaller surface unit cell we find that the vertical changes in position of the Zn and Ca atoms, compared to the position of
the Mg atom that they replace,
is $-0.34$ {\AA} (into the surface) for Zn and $+0.57$ {\AA} (out of the surface) for Ca, top of Table \ref{tab:zpos_combined}, 
in good agreement with the results of Ref.\ \cite{bolin26} in the larger surface cell.
Due to its larger atomic radius, the Ca atom does not fit into a Mg lattice, which explains why it is pushed out of the surface. 
In contrast, Zn has an atomic radius a bit smaller than that of Mg and therefore integrates more favorably into the Mg lattice
and is even dragged further in to the surface to have smaller distances to the second layer Mg atoms.  

\subsection{Adsorption of hydroxyapatite on alloyed Mg(0001)}

The adsorption energies of HA on pristine and alloyed Mg(0001) are summarized in the bottom part of Table \ref{tab:zpos_combined}. 
These show that the position of the alloying atom, whether Zn or Ca, matters for the stability of the system. 
The energetically most favorable position for Ca is (1,1), with adsorption energy of HA $-21.2$ meV/\AA$^2$,
whereas position (0,1) provides the most favorable configuration for Zn at $-22.4$ meV/\AA$^2$.
In general, the stability of both systems varies with the position of the substituted atom, and the adsorption energy 
can be either stronger or (in some cases) less strong than HA on a pure Mg surface, $-14.4$ meV/\AA$^2$. 

We calculate the atom positions in the interface between HA and the Mg surface after adsorption of a HA layer and compare
to (i) positions when HA is adsorbed on pristine Mg(0001), Table \ref{tab:distance_atoms}; and to (ii) positions of the alloying 
atom and the Mg atom positions before adsorption of a HA layer, Table \ref{tab:zpos_combined}.
From the first comparison we learn how the alloying affects the interface, and from the second comparison we learn how the 
Mg surface, whether pristine or alloyed, is affected by the adsorption of a HA layer.

\textit{Zn doping.}
Before adsorption of a HA layer, Zn takes a position in the Mg(0001) surface that is further  into the surface than that of 
the Mg atom it substitutes. With adsorption 
of a layer of HA on the already alloyed Mg surface, the largest change is that Zn at position (0,2)
under the lowest HA Ca atom moves 0.15 {\AA} further into the surface, Table \ref{tab:zpos_combined} row HA-Mg(Zn),
and ends up at a height $-0.49$ {\AA} relative to the Mg atom  in pristine Mg(0001)  without HA adsorption. 

\begin{figure}[tb]
    \centering
    \includegraphics[width=\linewidth]{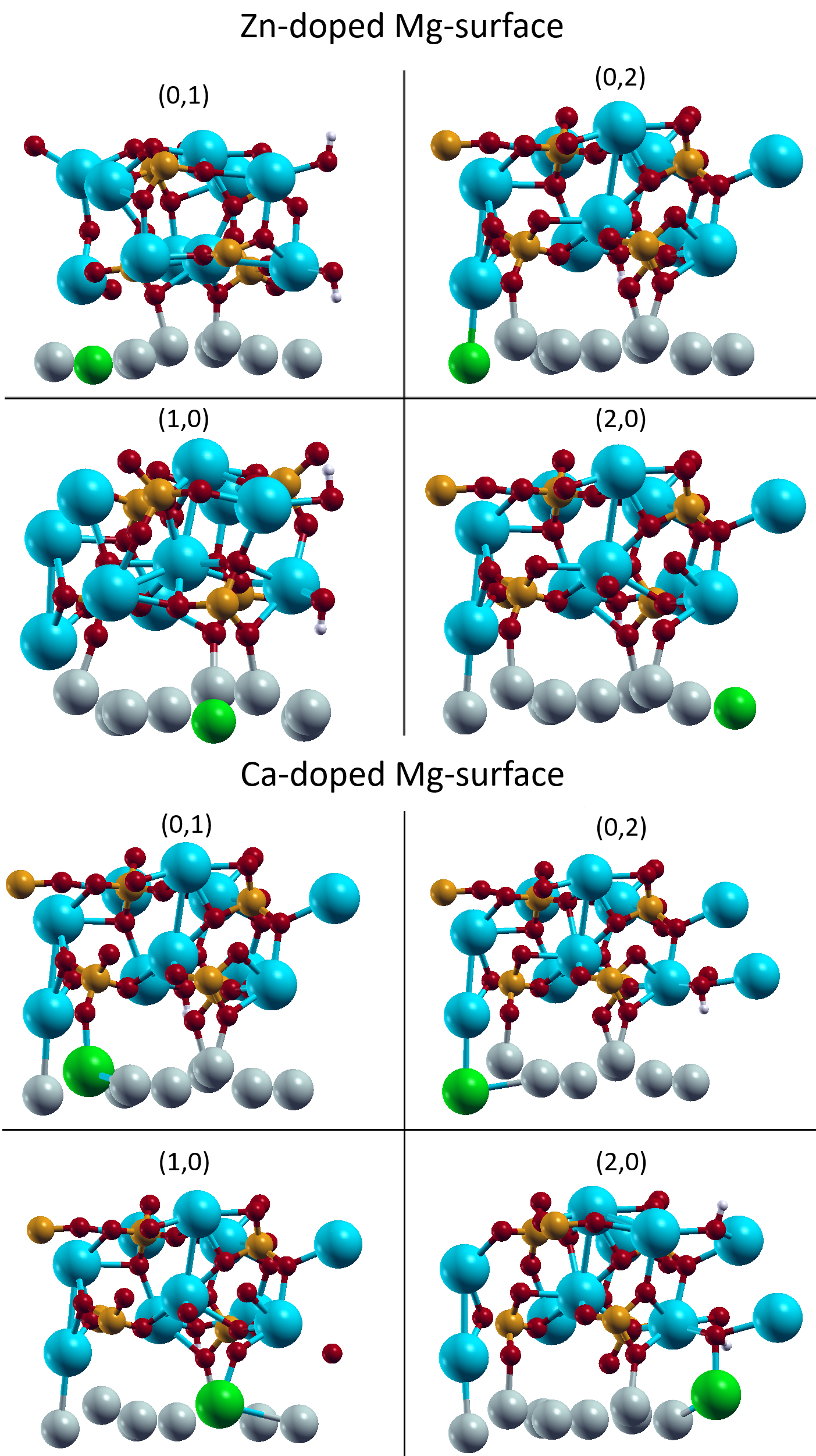}
    \caption{Optimized structures for Ca and Zn substitution at position (0,1), (0,2), (1,0) and (2,0). 
    Green atoms indicate substituted atom, in the top part a Zn atom, in the bottom part a Ca atom.
    Other atom colors as in Figure \protect\ref{fig:HA_mg_olika_vinklar}.}
    \label{fig:pos_zn_ca_efter_opt}
\end{figure}

The data for Zn alloying in Table \ref{tab:distance_atoms}, bottom part, shows that substituting Zn into the Mg surface
changes the relative atomic positions of Zn and the Ha atoms  in 
all directions, relative to pristine Mg(0001) with HA.
This is most pronounced for Zn in position (0,1), that is, under O$_\alpha$ before alloying, and for Zn under P, position (1,0);
compare also Figure \ref{fig:HA_before_after}(b) before alloying to Figure \ref{fig:pos_zn_ca_efter_opt} top panels (0,1) and (1,0) after alloying. 
There is clear repulsion of O from Zn, or rather, attraction to a Mg atom neighboring the Zn atom. 
This is because substitution with Zn in Mg(0001) changes the electron structure of
the neighboring Mg atom to become more attractive for O adsorption \cite{bolin26}. 

In general, however, the substitution of a Zn-atom in the top layer of Mg(0001) does not seem to have any major 
structural effect in HA or the surface. 

\begin{figure}[tb]
    \centering
    \includegraphics[width=\linewidth]{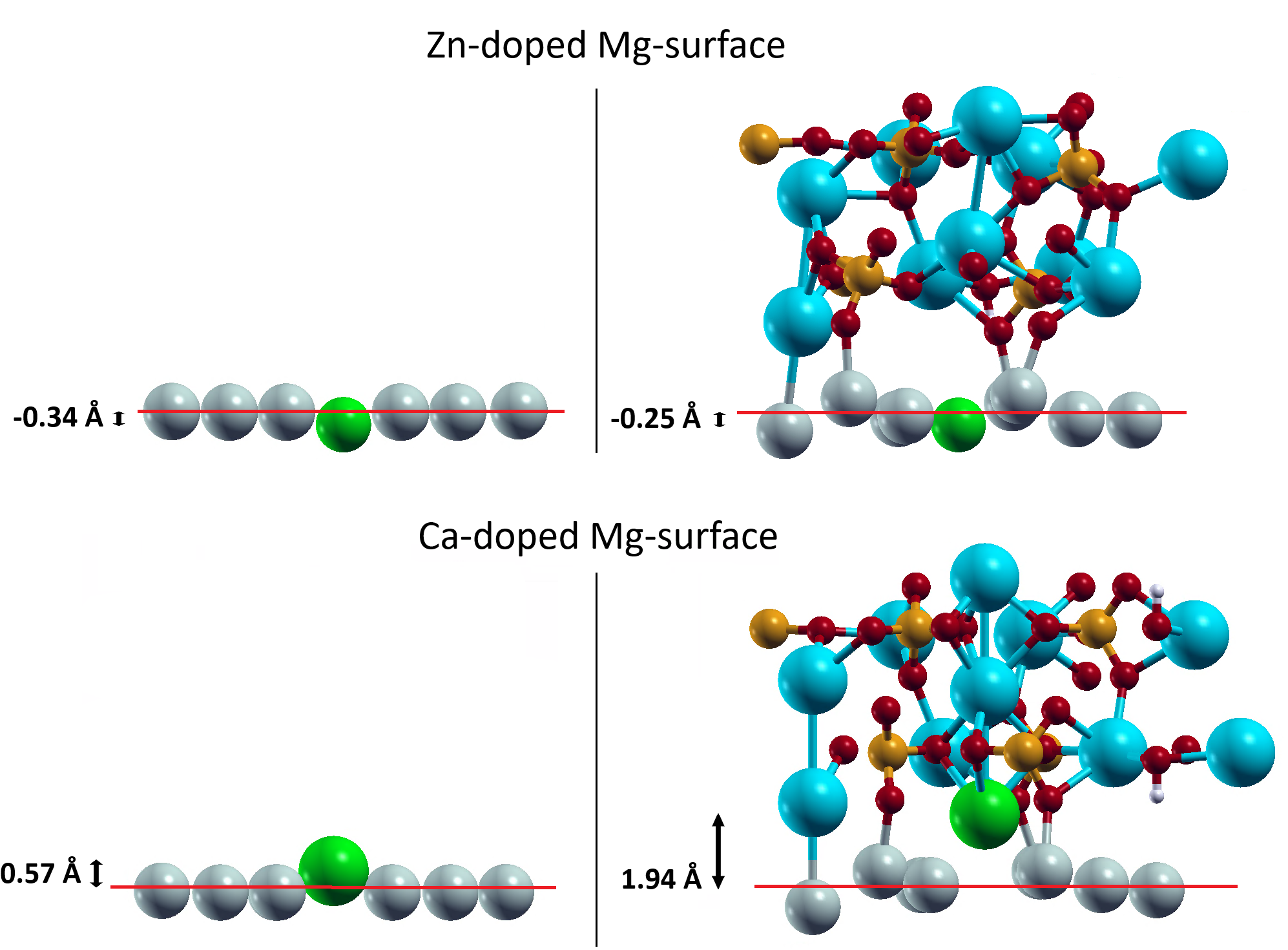}
    \caption{\textit{Left:} Mg(0001) with a doped atom substituted at position (1,1). 
    \textit{Right:} Same surface but relaxed with HA.
    Green atoms indicate substituted atoms. 
    The red reference lines indicate the position of Mg atoms in pristine Mg(0001) without HA.
    Atom colors as in Figure \protect\ref{fig:HA_mg_olika_vinklar}.
    }
    \label{fig:Ca_Zn_Movement}
\end{figure}

\textit{Ca doping.}
Introduction of a Ca alloying atom, compared to HA on pristine Mg(0001), shows a significant change when this
Ca atom is placed at position (1,1), Figure \ref{fig:Ca_Zn_Movement} bottom right.
Placed in that position, the Ca dopant moves further away from the Mg surface 
than in Mg(0001) without HA (compare Table \ref{tab:zpos_combined} rows Mg(Ca) and HA-Mg(Ca))
and into the HA structure.
Since Ca is one of the main atoms in HA, it is reasonable that a Ca atom in Mg(0001) tends to move closer to the HA layer, reflecting
its natural affinity for the HA structure. 
The alloy Ca atom enters the HA layer into a position similar to 
that of a Ca atom of neighboring layers in a HA bulk-like scenario. 
This position is also energetically favored, Table \ref{tab:zpos_combined} bottom.
We confirmed this scenario by a calculation of HA with three layers at initial separation 2 {\AA} in between each layer,
where the highest Ca atom in a lower layer was found to merge into the layer above, while remaining part of the HA layer,
rather than separating as observed on the Mg surface. 

Using Ca as a dopant at position (1,0) causes a vacancy in the Mg surface where the Ca atom 
leaves the surface. 

Generally, the Ca alloy atom tends to move away from the Mg surface, both relative to the position of Ca in Mg(0001) without adsorbed HA, 
and relative to the Mg atom it replaces in the HA-on-pristine-Mg situation.
This is likely due a combination of the larger size of the Ca atom compared to Mg, and the attraction to the O atoms in HA, and 
more generally the HA environment where Ca is already present.

The exception, with Ca moving into the surface instead of out, is the alloy Ca placed directly under
a low Ca atom in HA, position (0,2), where the two Ca atoms repel each other.
There, the alloy Ca atom is positioned at 0.57 {\AA} above the Mg surface when no HA is present, 
but changes to a position 0.80 {\AA} lower when HA is adsorbed, at vertical position 
$-0.80+0.57 =-0.23$ {\AA} (Table \ref{tab:zpos_combined}), 
i.e., the alloy Ca atom moves into the Mg surface due to the repulsion from the Ca atom in HA directly above.
The Ca-Ca distance (Table \ref{tab:distance_atoms}) is 3.186 {\AA}, which is closer than the Ca-Ca distance 3.95  {\AA}
in the stable face-centered cubic bulk structure; in the HA-on-Mg case the Ca atoms
are kept closer together by the other atoms around them.
This comes at a cost, energy wise: this position (0,2) has the weakest adsorption energy of all the studied 
situation, at $-5.2$ meV/{\AA}$^2$.

Nominally, in the position (2,0) the H-to-alloy-Ca distance increases, 
Table \ref{tab:distance_atoms} bottom part, compared to the case 
with pristine Mg(0001).
At the same time the alloy Ca atom moves further out of 
the Mg surface by 0.33 {\AA} (Table \ref{tab:zpos_combined}) once the HA is adsorbed.
This is because the O atom in the -OH group, of which the H atom is a part,
becomes available for binding to the alloy Ca, and this pushes the H atom out of the way, as seen 
in Fig.\ \ref{fig:pos_zn_ca_efter_opt} bottom, leading to a larger alloy-Ca-to-H distance.
Again, moving the H atom comes at a cost: the adsorption energy for the (2,0) 
situation is $-12.7$ meV/{\AA}$^2$, weaker than without the 
alloying and weaker than most of the alloying positions considered  here.

\begin{table}[tb]
\centering
\caption{Atom positional changes and adsorption energies of HA on Mg surfaces. 
Displacements are measured along the $z$-direction in \AA, adsorption energies in meV/\AA$^2$. 
The change in vertical position of Zn or Ca in Mg(0001) without HA, relative Mg atoms in pristine Mg(0001),
are given in the top part of the table, these are the reference points for displacements upon adsorption of HA 
(middle part).}
\label{tab:zpos_combined}
\vspace{0.2cm}

\begin{tabular}{l|ccccc}
\hline
Position & (0,1) & (0,2) & (1,0) & (1,1) & (2,0) \\
\hline
\multicolumn{6}{l}{\textit{No HA, displacement relative pristine Mg(0001)}} \\
\hline
Mg(Zn) & $-0.34$ & $-0.34$ & $-0.34$ & $-0.34$ & $-0.34$ \\
Mg(Ca) & 0.57 & 0.57 & 0.57 & 0.57 & 0.57 \\
\hline
\multicolumn{6}{l}{\textit{With HA, displacement relative to no HA}} \\
\hline
HA--Mg     & 0.54 & $-0.38$ & 0.63 & $-0.01$ & $-0.12$ \\
HA--Mg(Zn) & $-$0.03 & $-$0.15 & $-$0.06 & 0.09 & 0.06 \\
HA--Mg(Ca) & $-$0.01 & $-$0.80 & 0.08 & 1.37 & 0.33 \\
\hline
\multicolumn{6}{l}{\textit{Adsorption energy of HA on Mg(0001)}} \\
\hline
HA--Mg     & $-$14.4   & $-$14.4  & $-$14.4  & $-$14.4  & $-$14.4  \\
HA--Mg(Zn) & $-$22.4 & $-$19.8 & $-$14.2 & $-$15.3 & $-$16.5 \\
HA--Mg(Ca) & $-$20.0 & $-$5.2 & $-$20.7 & $-$21.2 & $-$12.7 \\
\hline
\end{tabular}
\end{table}

That the Ca atom in the energetically favorable position (1,1) moves out of the Mg surface and into HA
could have practical implications for the coating stability. 
While the differences in adsorption energies for various positions of the Ca alloy atom (from $-5$ to $-21$ meV/\AA$^2$) differ more than 
in the PES for pristine Mg in Figure \ref{fig:contour_graph} (10 meV/\AA$^2$ difference) 
HA may still be translated laterally at a low cost of energy under a relatively small applied force.
With Ca as an extra atom in HA micro-cracks may result, 
which would potentially compromise the integrity of the coating and its protective function.  

\subsection{Electron density}

\begin{figure}[tb]
    \centering
    \includegraphics[width=\linewidth]{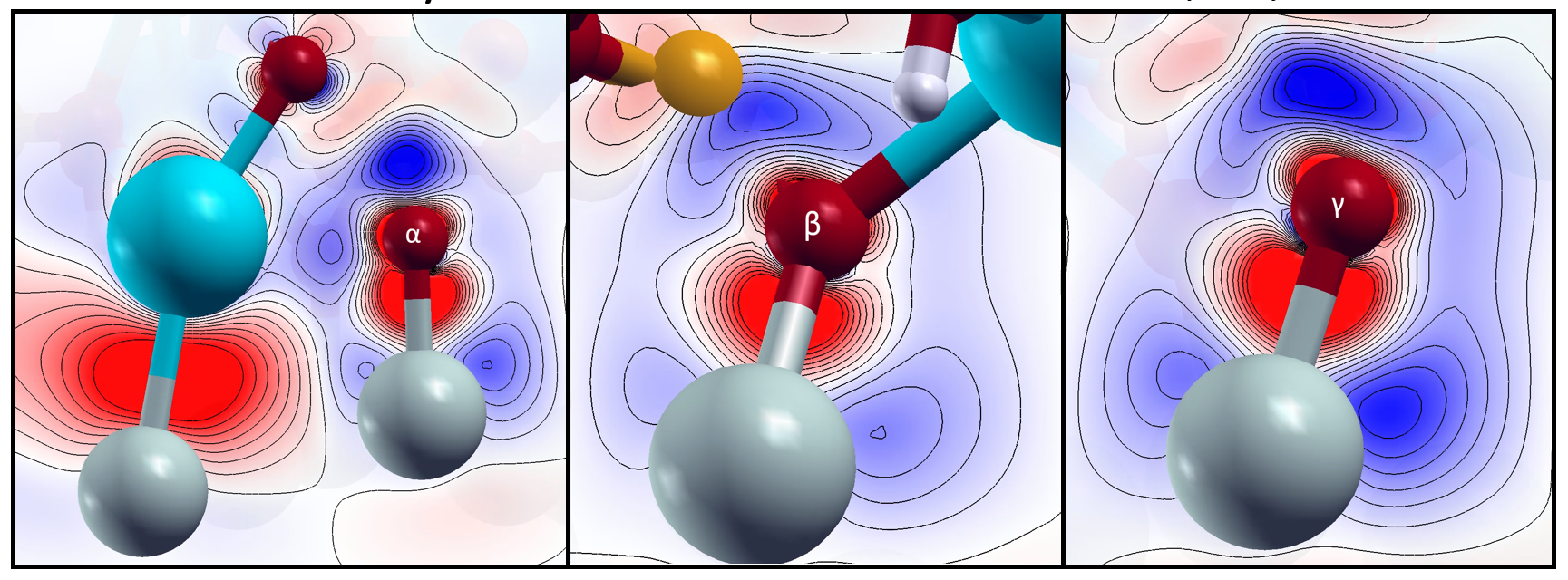}
    \caption{HA over pure Mg(0001); Change in electron density with red (blue) regions marking addition (depletion) of electron density, and isoline differences
   $7\cdot10^{-4}$ a.u. 
   Changes near the lowest Ca atom and the three O atoms marked $\alpha$, $\beta$ and $\gamma$ in Figure \protect\ref{fig:HA_mg_olika_vinklar}.
    Atom colors as in Figure \protect\ref{fig:HA_mg_olika_vinklar}.
    }
    \label{fig:Chargedens_full_HA}
\end{figure}

\begin{figure}[tb]
    \centering
    \includegraphics[width=\linewidth]{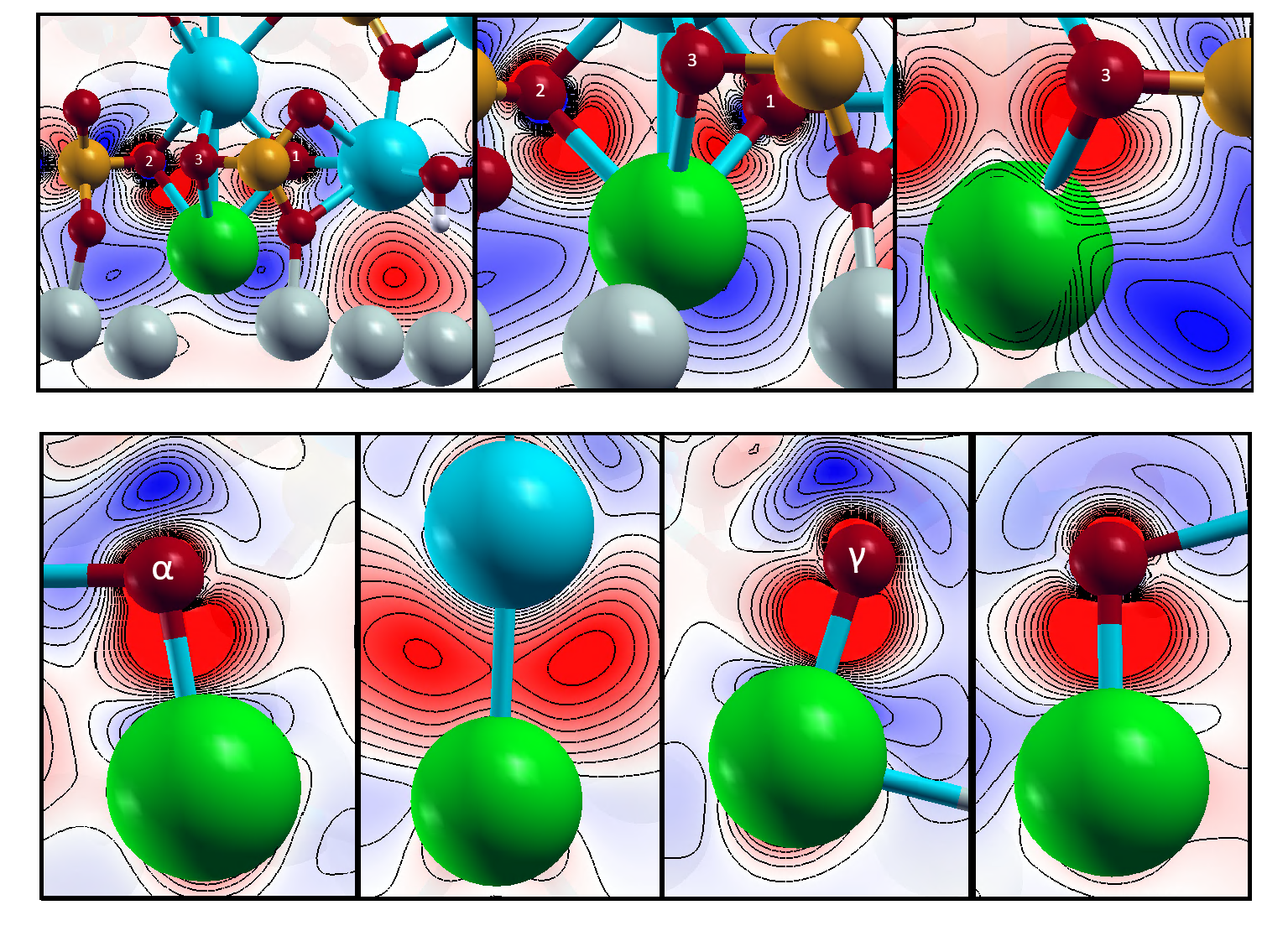}
    \caption{Change in electron density for HA adsorbed on the Ca-doped Mg surface. Isolines separated by $7\cdot10^{-4}$ a.u. 
    \textit{Top}: Ca atom at position (1,1), overview (left) and zoomed in on parts of the structure (middle and right).
    \textit{Bottom}: Ca in positions (0,1), (0,2), (1,0), and (2,0) (from left to right). 
    The H atom in (2,0) has moved and made the O atom in the -OH group 
    available for interaction with the Ca atom.
    Atom colors as in Figure \protect\ref{fig:HA_mg_olika_vinklar} and with the Ca alloying atom in green.
    }
    \label{fig:electron_density_all_ca_combined}
\end{figure}

\begin{figure}[tb]
    \centering
    \includegraphics[width=\linewidth]{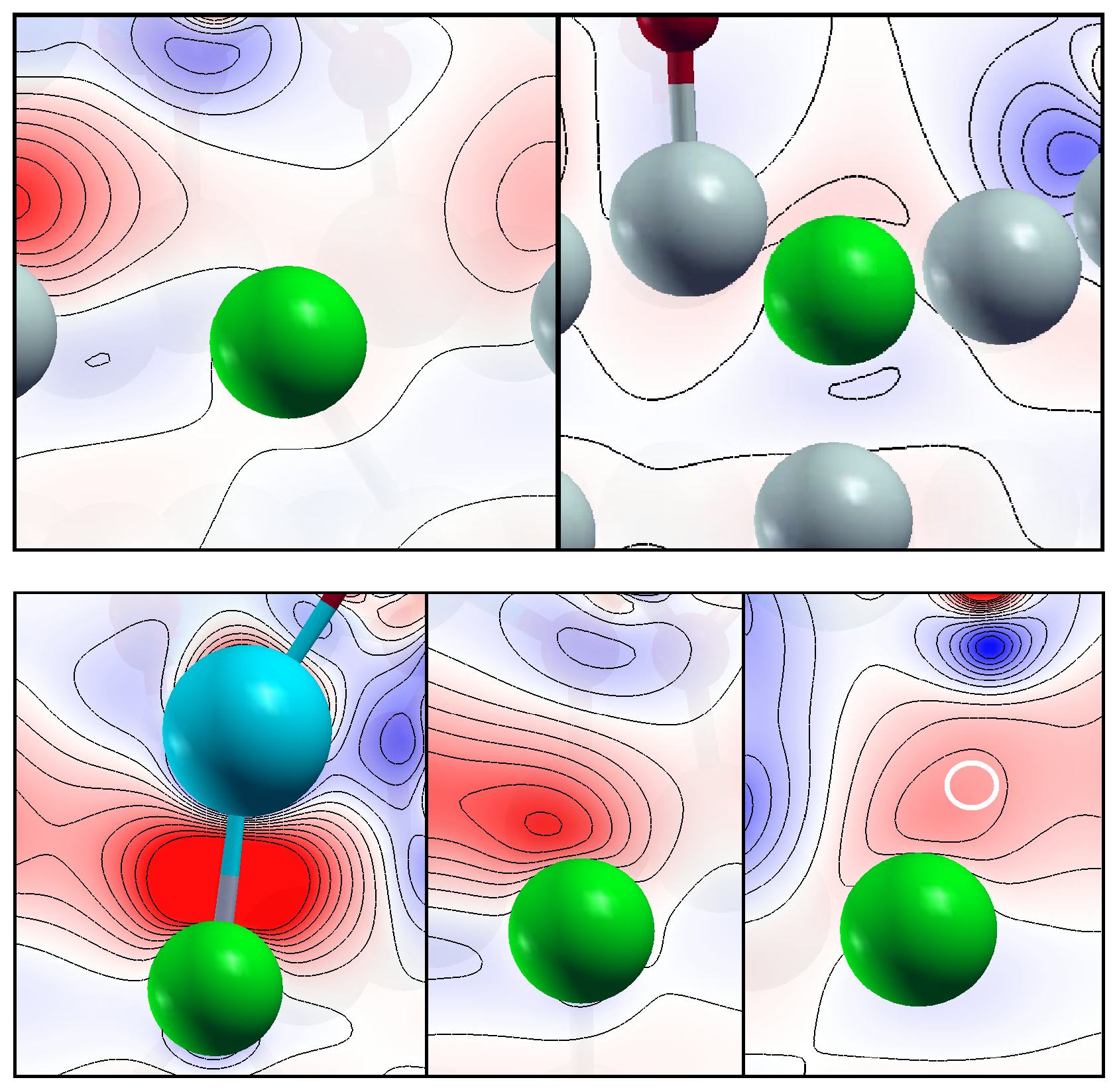}
    \caption{Change in electron density for HA adsorbed on the Zn-doped Mg surface. Isolines separated by  $7\cdot10^{-4}$ a.u.
    \textit{Top}: Zn atom at position (0,1) (left) and (1,1) (right);
    \textit{Bottom}: 
    Zn atom at positions (0,2), (1,0), and (2,0) (left to right). 
    In the panel for (2,0) a H atom is slightly behind the plane of the panel and its position is marked by a white circle.
    Atom colors as in Figure \protect\ref{fig:HA_mg_olika_vinklar}, and Zn in green.
    }
    \label{fig:chargedens_all_zn_complete}
\end{figure}

To better understand the interaction between HA and the Mg surface, electron density plots were generated for the optimized structures, 
analyzing the interface for both pure and doped Mg(0001). 
Changes in the electron density are evaluated similarly to the adsorption energy, but without any additional structural relaxation calculations
when HA and surface are pulled apart: 
The electron density of the combined system is compared to the sum of the electron densities of the isolated but still deformed
HA layer and  Mg surface. 
In Figures \ref{fig:Chargedens_full_HA}-\ref{fig:chargedens_all_zn_complete} we thus plot the electron density difference 
due to the adsorption of the HA layer.
Isolines have separation $7\cdot10^{-4}$ a.u.\ (electrons/Bohr$^3$), and the color scale goes from 
$5\cdot10^{-3}$ a.u.\ (red, increased electron density) to $-5\cdot10^{-3}$ a.u.\ (blue, decreased electron density).

Changes in electron density for HA adsorption on pristine Mg is visualized in Figure \ref{fig:Chargedens_full_HA}. 
The largest electron accumulations are near the three O-atoms (O$_\alpha$, O$_\beta$ and O$_\gamma$) that are close to the Mg atoms,
and between the lowest Ca atom in HA and a Mg atom. 
Meanwhile, electron depletion is also seen around the Mg atoms and above the O atoms in the directions to other atoms.

Figure \ref{fig:electron_density_all_ca_combined} shows the Ca-doped system, with focus on the positions of the Ca alloying atom in the five 
positions examined, Fig.~\ref{fig:Sub_atoms}(b).
With Ca in position (1,1), top panels of Fig.~\ref{fig:electron_density_all_ca_combined}, 
the maximum electron density change is $3.6\cdot10^{-2}$ a.u. This is larger than for any of the other systems. 
We know from the structural changes illustrated in Figure \ref{fig:Ca_Zn_Movement} that when Ca is placed in position (1,1) it moves into the HA layer
to a spot surrounded by three O atoms (marked by numbers 1, 2, and 3 
in Figure \ref{fig:electron_density_all_ca_combined}). 
These O atoms are positioned further into the HA layer than the O atoms marked by $\alpha$, $\beta$ and $\gamma$ in previous figures.
Given the high reactivity of O and its strong tendency to attract cations, it is reasonable that Ca stabilizes in a position close 
to these O atoms and consistent with the electron accumulation between the alloy Ca atom and the O atoms with labels 1, 2, and 3.
 
The other examined interfaces of Ha with Ca-doped Mg(0001) are shown in the bottom panels of Figure \ref{fig:electron_density_all_ca_combined}.
The electron accumulation between the alloy Ca atom and the HA Ca atom, position (0,2), is more delocalized compared to the interaction between Ca and O. 
This is to be expected because Ca atoms have low electronegativity, neither of the Ca atoms attract the excess electron density strongly to localize it 
between them, leading to two lobes on each side instead of a single lobe in the middle. 
Meanwhile, the alloy Ca atom and O in HA shows a typical interaction between an electronegative and electropositive atom in the three other bottom panels,
for Ca alloy positions (0,1), (1,0) and (2,0).
Thus, O strongly attracts electron density whereas Ca donates electron density from its outer shell, leading to the localized lobe 
rather than the more delocalized pattern observed between the two Ca atoms. 

In the HA interfaces with the Zn-doped surfaces, Figure \ref{fig:chargedens_all_zn_complete}, the electron density changes 
are smaller than in the other systems.
As seen in the middle row of Table \ref{tab:zpos_combined} the
Zn-atom does not move much when HA is adsorbed on the already alloyed Mg surface.  
This is reflected in the electron density changes: 
There is very little electron accumulation between Zn and neighboring atoms in position (0,1) (O neighboring atom) and (1,1) (mostly Mg atoms nearby), and
in position (0,2) the Zn to Ca interaction shows the largest electron accumulation among the Zn-doped systems, but still smaller 
than in most of the systems with Ca-alloyed or pristine Mg(0001). 
In position (1,0) the Zn atom has P, Ca and some O atoms as nearest neighbors but at a distance such that the electron changes
that these give rise to are spread out. 
Finally, in position (2,0) the H atom above Zn is out of the vertical plane through Zn and 
the position of the H atom is indicated by the white circle in the panel. There, we see some low level electron accumulation.

The electron density change plots overall show larger variation for HA on Ca doped surfaces than for HA on Zn doped surfaces.
This is in agreement with the differences in adsorption energy being larger for HA on Ca doped surfaces than on Zn doped surfaces. 

\section{Summary}

We investigated the interaction of one layer of HA with pristine and doped Mg(0001), where for the doping we used one Zn or Ca atom 
to replace one out of 18 exposed Mg atoms in a $3\times3$ Mg(0001) surface cell.
Our calculations show that HA binds to the Mg surface but that the interaction is relatively weak, 
with an adsorption energy of $-14.4$ meV/\AA$^2$ on pristine Mg(0001) and between $-5.2$ and $-22.4$ meV/\AA$^2$ on doped Mg(0001), 
depending also on the relative position of the dopant and the HA layer. 
For HA on pristine Mg(0001) we studied the corrugation by calculating the adsorption energy at 
positions distributed on the surface, and found that also the corrugation is low, with 9.9  meV/\AA$^2$ as the difference between 
the energetically worst and best adsorption position. 
Thus the energy cost of HA sliding across the surface is low. 

Doping the Mg surface with Ca or Zn affects the interaction with HA. 
Ca-doped surfaces show a variation of stability and deformations, with the best adsorption energy in a situation where the Ca 
atom leaves the Mg surface and enters the HA layer.
Zn-doped surfaces do not show any major deformations at HA adsorption, 
but still have better adsorption energy for HA than the pristine Mg surface.

Plots of the changes in electron density upon adsorption of HA show that the changes occur mainly between the lowest O atoms of HA
and the Mg surface or the Ca atom of the doped surface, with electron accumulation near the O atoms,
while there is less electron accumulation between the O atoms and Zn. 
Further, we find relatively large electron density changes between the lowest Ca atom of HA and the Mg, Ca or Zn atom below it.

\section*{Author contributions (CR{edi}T)}
\textbf{Anthony Veit Berg:} Conceptualization (equal), Formal analysis (supporting), Investigation (lead), Visualization, Writing - original draft (equal).
\textbf{Ablai Forster, Tim Hansson, Alexandra J. Jernstedt, and Emmy Salminen:} Conceptualization (equal), Investigation (supporting), Writing - review \& editing.
\textbf{Elsebeth Schr\"oder:}  Conceptualization (equal), Data curration, Formal analysis (lead), Funding acquisition, Investigation (supporting), Resources, Supervision, Validation, Writing - original draft (equal).

A.F., T.H., A.J., and E.S.\ (Emmy Salminen) contributed equally to this work. 

\section*{Acknowledgment}
The present work is supported by the Swedish Research Council (VR) through Grant No.~2020-04997 and Chalmers Areas of Advance Nano and Materials.
The computations were performed using computational and storage allocations from
Chalmers Centre for Computational Science and Engineering (C3SE), 
and from the 
National Academic Infrastructure for Supercomputing in Sweden (NAISS), under contracts
C3SE2025/1-6,   
C3SE2026/1-21, 
C3SE2026/1-23, 
NAISS2024/3-16, 
NAISS2024/6-432, 
NAISS2025/3-25, and 
NAISS2025/5-484. 

The authors have no conflicts of interests to disclose.

\end{document}